\documentclass[english]{sbrt}
\usepackage[english]{babel}
\usepackage[utf8]{inputenc}

\usepackage{graphicx}
\usepackage{subfig}
\usepackage{amsmath}
\usepackage{amssymb}

\begin{document}

\title{Physical Layer Security Techniques Applied to Vehicle-to-Everything Networks}

\author{Leonardo B. da Silva, Evelio M. G. Fernández and Ândrei Camponogara \thanks{L. B. da Silva, E. M. G. Fernandez, Â. Camponogara, Electrical Engineering Department, Federal University of Paraná (UFPR), Curitiba, PR, Brazil, e-mails: leonardobarbosa@ufpr.br, evelio@ufpr.br and andrei.camponogara@ufpr.br. This study was financed in part by the Coordenação de Aperfeiçoamento de Pessoal de Nível Superior – Brasil (CAPES) – Finance Code 001.}%
}

\maketitle

\begin{abstract}
Physical Layer Security (PLS) is an emerging concept in the field of secrecy for wireless communications that can be used alongside cryptography to prevent unauthorized devices from eavesdropping a legitimate transmission. It offers low computational cost and overhead by injecting an interfering signal in the wiretap channels of potential eavesdroppers. This paper discusses the benefits of the Artificial Noise and Cooperative Jamming techniques in the context of Vehicle-to-everything (V2X) networks, which require secure data exchange with small latency. The simulations indicate that messages can be safely delivered even with devices that have low available power.
\end{abstract}

\section{Introduction}
\label{section: intro}

Urban mobility is one of the main focuses of the Internet of Things (IoT) when applied to smart cities, due to the necessity for more responsive and safe traffic control. Generally, the solutions proposed in this scope involve the wireless communication between not only the vehicles themselves, but also with pedestrians, infrastructure, and networks. This paradigm is known as Vehicle-to-everything (V2X) and it can be standardized by protocols such as C-ITS (Cellular Intelligent Transportation System) and WAVE (Wireless Access for Vehicular Environment) that are based on the IEEE 802.11p amendment, and the Cellular-V2X (C-V2X) that implements the 5G standard from 3GPP (3rd Generation Partnership Project) \cite{ElHalawany2019}. 

\subsection{Problem Outline}

Due to the ever-changing location of most of the involved communication nodes and the time-sensitive nature of the data involved (brake position, vehicle speed, traffic volume, accident reports, etc), the transmission needs not only to occur at high rates, but also offer reliability through high secrecy, low packet loss, and small delay. Furthermore, those nodes have to be affordable to justify their implementation on a city-wide scale, thus having low power consumption and the most cost-efficient embedded processing unit possible \cite{Hamamreh2019}.    

Since the main source of information security in today's landscape is provided through cryptography, the secrecy constraint can negatively affect most of these criteria. As a result of the growth in the availability of portable and connected equipment with high processing capabilities, the safety measures implemented need to match this computational power with proportionally longer and more complex keys to not be vulnerable to brute-force attacks from well-equipped malicious devices \cite{Hamamreh2019,Sanenga2020}. This approach, however, is not sustainable, because it produces increasingly long authentication routines, due to the raise in computational overhead and processing cost as a result of the implemented security algorithms.

\subsection{Overview of the proposed solution}

To counterbalance this issue, this paper studies the use of Physical Layer Security (PLS) techniques as an additional protection to increase the secrecy of wireless communications in a V2X environment. As the name suggests, PLS is applied at the Physical Layer, making it an alternative that can be used with low processing cost when compared with cryptography, which is more oriented towards the computational side of the network stack on the Application Layer \cite{ElHalawany2019}. 

Since cryptography techniques provide security in different sections of the wireless protocols, PLS is proposed as a complement to them, rather than a replacement \cite{ElHalawany2019}. Through the use of both approaches on the same node, it is possible to offer high secrecy without the necessity of infinitely growing key complexity. 

The PLS has its origins on the analytical proposal of Wyner's wiretap channel \cite{Wyner1975}, where it is described a communication between two legitimate nodes that is spied on by an eavesdropper through an unauthorized channel called wiretap. In the modern literature, these devices are usually referred to as a transmitter called Alice, an authorized receiver Bob, and the set of $K$ eavesdroppers named Eves.

\begin{figure}[hbt]
    \begin{center}
        \setlength{\unitlength}{0.0105in}%
        \includegraphics[width=\linewidth]{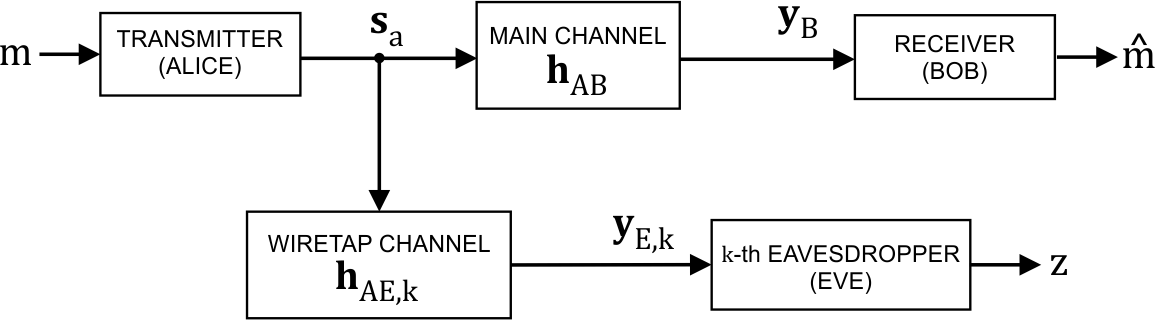}
    \end{center}
    \caption{\label{fig: wiretap_channel} The wiretap channel generic model based on \cite{Wyner1975}}
\end{figure}

In the wiretap channel model shown in Fig. \ref{fig: wiretap_channel}, the original message $m$ is encoded and transmitted by Alice as the signal $\textbf{s}_a$, that reaches Bob through the main channel $\textbf{h}_{AB}$. The received signal $\textbf{y}_B$ is then decoded by Bob, obtaining the estimated message $\hat{m}$. Additionally, the $k$-th Eve can intercept $\textbf{s}_a$ through the wiretap channel $\textbf{h}_{AE,k}$, obtaining the signal $\textbf{y}_{E,k}$ that when decoded produces $z$. 

The main focus of PLS is to guarantee that the mutual information between $m$ and $z$ is as close to zero as possible. When this condition is met, even if $z$ is know, it is impossible for Eve to infer the contents of the original message.

Wyner then presents a set of parameters that enable the use of the physical imperfections of the channel, such as noise and fading, to provide information secrecy by raising the level of confusion on undesired nodes. Rendering them unable to distinguish between the message and the interference.

Currently, plenty of techniques to provide security at the physical layer level have been proposed in the literature \cite{Sanenga2020}. This paper will focus on two approaches first presented in \cite{Negi2005}:

\vspace{.1em}
\begin{itemize}
    \item \textbf{Artificial Noise (AN):} This approach uses a portion of the transmitter node's power to inject artificially generated noise in the eavesdropper's channel;
    \item \textbf{Cooperative Jamming (CJ):} This approach expands the AN model by proposing a connected network where nearby relay nodes (Charlies) send a jamming signal to the eavesdropper's channel.
\end{itemize}
\vspace{.1em}

To demonstrate the viability of AN and CJ applications in a V2X network, it is common to create stochastic geometric models that randomly generate streets and distribute communication nodes in a predefined area to represent an urban mobility scenario \cite{Wang2020,Qiu2020}. When implementing these methods, metrics such as the Signal-to-Interference Ratio (SIR) are used to define the threshold of confusion necessary to provide secrecy at the physical layer. The SIR on each eavesdropper can then be evaluated to determine the secrecy outage probability (SOP) of the data transmission with different densities of the involved nodes in the simulated network. 

In this paper, Section \ref{section: V2X_net} describes the stochastic algorithms implemented to model a V2X network that includes streets and communication nodes (vehicular and planar). Section \ref{section: PLS_techs} presents the analytical basis of the AN and CJ techniques, while also introducing the SIR and SOP metrics. In Section \ref{section: Num_sims}, the results of numerical simulations are shown to illustrate the benefits of the considered PLS techniques on the generated V2X networks. Finally, Section \ref{section: conclusion} states some final remarks. 

\textit{Notation}: $\textbf{I}_{N}$ is an identity matrix of order $N$, Poisson($n$) is a Poisson distribution with mean number of arrivals $n$, $\mathcal{CN}(m,n)$ is a complex normal distribution with average $m$ and covariance $n$, $\exp(n)$ is an exponential distribution with mean $n$ and Gamma$(m,n)$ is the gamma distribution with form $m$ and scale $n$.

\section{The V2X network model}
\label{section: V2X_net}

As mentioned previously, vehicular networks are dynamic, with devices changing location constantly. Thus, a deterministic model is not well-suited for this application. A common alternative is the use of stochastic geometry to represent this random spatial nature through a variety of different processes to distribute the streets and communication nodes within the desired coverage area \cite{Haenggi2012}.   

\begin{figure}[hbt]
    \begin{center}
        \setlength{\unitlength}{0.0105in}%
        \includegraphics[width=\linewidth]{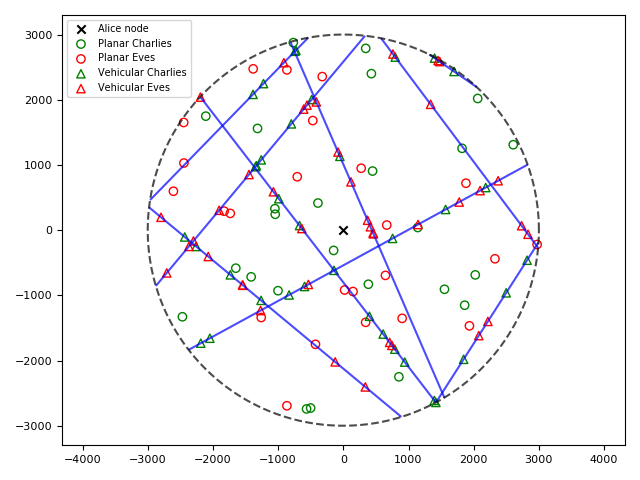}
    \end{center}
    \caption{\label{fig: spatial_sim} Spatial simulation of the modeled V2X network. The color green indicates the Charlies implemented in CJ techniques and the Eves are in red. The planar devices are generated by PPPs represented by circles ($\circ$) with intensity $\lambda$ = 10$^{-6}$/m$^2$ for both node types. Through a PLP, the streets (blue lines) have been modeled with an intensity of $\lambda_l$ = 10$^{-3}$ /m, and the vehicular devices are originated from PLP-driven Cox Processes indicated with triangles ($\triangle$) of intensity $u$ = 10$^{-3}$/m for both Charlies and Eves. A single Alice is indicated with a black $\times$ at the origin.}
\end{figure}

A viable option is the use of Poisson processes, as they are memoryless counting processes for integer arrivals \cite{Yates2005}. In other words, each set of elements generated will be independent with a Poisson distributed integer number of uniformly spaced nodes. The intensity of the arrivals in these processes are represented by $\lambda$ and the expected number of elements is the product of the said intensity and the Lebesgue measure, which in this context is essentially the spatial measurement associated with the object that the points will be distributed on. For instance, the Lebesgue measure to populate a circle is its area and for a line is the length. One realization of the resulting spatial model derived from the use of different variations of the Poisson processes is represented in Fig. \ref{fig: spatial_sim}.

In this model, the wireless devices of pedestrians and connected infrastructure are considered free to be positioned in the whole area $A$ of the modeled network, which is a circle of radius $r$ = 3 km. Thus, these ``planar nodes`` are generated by 2-D Poisson Point Processes (PPP) and the expected amount of elements is given by Poisson($\lambda \cdot A$). The set of planar nodes is indicated by $\Phi$, thus the planar Eves and Charlies are respectively represented by $\Phi_E$ and $\Phi_C$.

The streets are represented by uniformly distributed lines with density $\mu_l = \lambda_l/\pi$ generated by a Poisson Line Process (PLP) $\Phi_l$ based on the second method of the Bertrand paradox \cite{Bertrand1889}, in which a set of expected Poisson($\mu_l \cdot 2\pi r$) midpoints are created \cite{Chetlur2018}, each with a random radius $P \in$ [0, $r$) and angle $\theta \in [0, 2\pi)$. From these coordinates, a segment perpendicular to $P$ is traced between two points at the edge of the circle of radius $r$. This effectively means that a pair of 1-D PPP points are created in the perimeter of the circular area for each modeled street. 

On those PLP-generated lines, a Cox process of intensity $u$ is implemented, which is used to create the ``vehicular nodes`` on each segment \cite{Choi2018}. These elements represent vehicles whose spatial distribution are constrained to a street by a 1-D PPP. Considering a street of length $l$, the number of vehicles in it is given by Poisson($u \cdot l$). 

The set of vehicular Eves and Charlies on each street $l$ are respectively denoted by $\psi_E$ and $\psi_C$. Based on these, the total nodes of each type can be obtained by evaluating the sets on the whole range of $\Phi_l$ \cite{Wang2020}, resulting in $\Psi_E$ = $\{ \psi_E(l) \}_{l\in\Phi_{l}}$ for Eves and $\Psi_C$ = $\{ \psi_C(l) \}_{l\in\Phi_{l}}$ for Charlies.

Furthermore, a single deterministic transmitter (Alice) is included at the origin of the circle. This point is selected to simplify the distance calculations between a legitimate device and the Eve nodes, which can be planar or vehicular. This measurement is one of the parameters for the SIR calculations, that are considered to determine the effectiveness of the PLS. For the CJ case, auxiliary nodes (Charlies) are also modeled, some as planar and others as vehicular devices. Note that the distance between Charlies and Eves influences the power of the interference injected on the unauthorized channels as part of the jamming technique.

\section{PLS Techniques}
\label{section: PLS_techs}

The PLS techniques presented in this paper are part of the key-less-based class \cite{Hamamreh2019}, which implements secure information transmission by making the unauthorized channel's capacity ($C_E$) lower than that of the legitimate channel's ($C_B$). This relationship can be presented by evaluating these values through the Shannon-Hartley theorem, which produces the secrecy capacity ($C_S$) metric as
\begin{equation}
    C_S = C_B - C_E = \log_2(1+ \gamma_B) - \log_2(1+ \gamma_E),
    \label{eq: C_S}
\end{equation}

\noindent where $\gamma_B$ and $\gamma_E$ are, respectively, the SIRs of Bob and Eve. Based on this expression, it can be inferred that in order to guarantee that $C_B$ is sufficiently larger than $C_E$, the  value of $\gamma_E$ must be as low as possible. The approach utilized by AN and CJ is the injection of artificially generated interference in the eavesdropper channels.

Typically, this injection is implemented with multi-antenna networks, as it enables the use of beamforming to selectively direct the transmission to legitimate receivers with minimum noise and high efficiency \cite{Sanenga2020}. The unintended receivers on the other hand, intercept a signal that contains the secret message as well as AN. Therefore, secrecy is provided when the distinction between them by the Eves is improbable.

The wireless channels in this paper are modeled with complex normal distributions ($\mathcal{CN}$) which implies in a Rayleigh fading model. This decision provides simpler analytical equations and also proposes a more pessimistic scenario, in which there is no Line-of-Sight (LoS) available. By evaluating the metrics in these worst-case conditions, it is possible to verify that even then the secrecy can be guaranteed.  

\subsection{Artificial Noise}

In the AN scenario, the legitimate communication is established between a single transmitter Alice and a receiver Bob. Additional nodes (both planar and vehicular) that try to obtain Alice's signal are then considered eavesdroppers and their channels will be affected by the AN.

The signal transmitted by the Alice node with $N_A$ antennas is composed of two terms: the first contains a message $x$ intended for Bob and the second is based on a zero-forcing vector for the unauthorized devices \cite{Hu2018}, i.e,

\begin{equation}
    \textbf{s}_a = \sqrt{\phi P_t} \frac{\textbf{h}_{a}}{\|\textbf{h}_{a}\|}x + \sqrt{\frac{(1-\phi)P_t}{N_A - 1}}\textbf{W}_a \textbf{n}_a,
    \label{eq: s_a}
\end{equation}

\noindent where $\textbf{h}_{a}/\|\textbf{h}_{a}\|$ is the beamforming vector with the normalization of the Alice's channel estimation $\textbf{h}_{a} \in \mathbb{C}^{N_A \times 1}$, that will be modeled as $\mathcal{CN}(0,\textbf{I}_{N_A})$. The AN is formed by the null-space orthonormal basis $\textbf{W}_a \in \mathbb{C}^{N_A \times (N_A - 1)}$ and the noise signal $\textbf{n}_a \in \mathbb{C}^{(N_A - 1) \times 1}$. 

The distribution of the available power, $P_t$, between the two terms of (\ref{eq: s_a}) is controlled by $\phi \in$ \{0,1\}. $\phi = 0$ means that all power is allocated to noise generation and no message is sent. Conversely, when $\phi$ = 1 the AN is not active and $P_t$ is allocated entirely for data transmission.

\subsection{Cooperative Jamming}

The Cooperative Jamming extends the AN case, maintaining the single Alice-Bob authorized transmission with multiple Eves, however, adding auxiliary nodes in the network. These devices, typically called Charlies, can also be either planar or vehicular, just like the Eves. In contrast, they are responsible for providing additional security by sending jamming signals that further decrease the channel quality of the Eves.

For simplicity, it is considered that only Alice will transmit messages in the scenarios evaluated in this paper. Hence, the signals sent by the Charlie nodes are made of only the AN (zero-forcing) portion, as follows

 \begin{equation}
     \textbf{s}_c = \sqrt{\frac{P_c}{N_C-1}} \textbf{W}_c \textbf{n}_c,
     \label{eq: s_c}
 \end{equation}

\noindent where $N_C$ is the number of antennas of each Charlie and $P_C$ is the power available for jamming. Notice that since these nodes are not transmitting messages, all the available power is directed towards CJ. Additionally, $\textbf{W}_c \in \mathbb{C}^{N_C \times (N_C - 1)}$ is the null space orthonormal matrix and $\textbf{n}_c \in \mathbb{C}^{(N_C - 1) \times 1}$ is the artificial noise component.

\subsection{Received Signals}

By considering that the channel estimation $\textbf{h}_{a}$ is precisely the main channel established between Alice and Bob, $\textbf{h}_{AB}$, it is implied that the receiver node is not affected by the interference from AN or CJ. That happens because the orthonormal basis $\textbf{W}_a$ and $\textbf{W}_c$ are null when applied to the authorized channels, resulting in the relationships $\textbf{h}_{AB}^{\dag} \textbf{W}_a = 0$ and $\textbf{h}_{AB}^{\dag} \textbf{W}_c = 0$, respectively. Therefore, the signal received by Bob can be expressed as

\begin{equation}
    \textbf{y}_B = \: \sqrt{\phi P_t} \: \|\textbf{h}_a\| \: D_{AB}^{-\alpha/2} \: x,
    \label{eq: y_B}
\end{equation}

\noindent
where $D_{AB}$ is the distance between the devices and $\alpha > 2$ is the path loss exponent considering an NLoS scenario. The distances are obtained through simple trigonometry based on the coordinates randomly generated by the stochastic processes described in Section \ref{section: V2X_net}.

For the signal intercepted by the eavesdroppers, it is evaluated a set of $K = (\Phi_E + \Psi_E)$ Eves, containing both planar and vehicular nodes. Similar considerations are adopted for the Charlies in the CJ scenario, resulting in $C = (\Phi_C + \Psi_C)$.

As discussed when $\textbf{s}_a$ was presented, Alice sends a signal containing the secret information and AN. Since authorized Alice-Eves channels are not expected in the beamforming sense, the orthonormal basis are not null, thus the Eves receive interference. When the Cooperative Jamming is taken into consideration, Eves are also affected by the interference generated by the nearby Charlies through the $\textbf{s}_c$ signals. With that in mind, the signal obtained by the $k$-th Eve is given by 
\begin{align}
    \begin{split}
        \textbf{y}_{E,k} &= \sqrt{\phi P_t} \: \textbf{h}_{AE,k}^{\dag} \: D_{AE,k}^{-\alpha/2} \: x \\
        &+ \sqrt{\frac{(1-\phi)P_t}{N_A-1}} \: \textbf{h}_{AE,k}^{\dag} \: \textbf{W}_a \: D_{AE,k}^{-\alpha/2} \: \textbf{n}_a  \\
        &+ \sum\limits_{c \: \in C}\sqrt{\frac{P_c}{N_C-1}} \: \textbf{h}_{c,k}^{\dag} \: \textbf{W}_c \: D_{c,k}^{-\alpha/2} \: \textbf{n}_c \:, 
        \label{eq: y_Ek}
    \end{split}
\end{align} 

\noindent which is composed of essentially three terms. The first is the intercepted secret message itself, the second term is the AN signal generated by Alice, and the third term is a sum of all the interference injected by the Charlie nodes. Since CJ only affects the last term of (\ref{eq: y_Ek}), the AN scenario can be obtained by simply adopting that the sum in this term is equal to zero.

From (\ref{eq: y_B}) and (\ref{eq: y_Ek}), it is possible to determine the SIR of Bob and the $K$ Eves. Thus, the SIR of Bob can be determined as

\begin{equation}
    \gamma_{B} = P_t\phi \: \left\|\textbf{h}_{a}\right\|^2 D_{AB}^{-\alpha},
    \label{eq: SIR_B}
\end{equation}

\noindent and the SIR for each Eve can be obtained from (\ref{eq: y_Ek}) as follows

\begin{equation}
    \gamma_{E,k} = \dfrac{P_t \: \phi \: \left|\textbf{h}_{AE,k}^{\dag} \: \large {\textbf{h}_{a}/\|\textbf{h}_{a}\|}\right|^2 D_{AE,k}^{-\alpha}}{\frac{P_t \: (1-\phi)}{N_A-1} \: \left\|\textbf{h}_{AE,k}^{\dag} \: \textbf{W}_a\right\|^2 D_{AE,k}^{-\alpha} + I_c} , 
    \label{eq: SIR_Ek}
\end{equation}

\noindent where $I_c$ is the sum of the interference injected by the Charlies given by

\begin{equation}
    I_c = \sum_{c \: \in \: C}\frac{P_c}{N_c-1}\|\textbf{h}_{c,k}^{\dag} \: \textbf{W}_c\|^2 \: D_{ck}^{-\alpha},
\end{equation}

\noindent which is non-zero only in the CJ scenario. The products $\textbf{h}_{AE,k}^{\dag} \cdot {\textbf{h}_a/\|\textbf{h}_a\|}$ and $\textbf{h}_{AE,k}^{\dag} \cdot \textbf{W}_a$ from the Alice-Eve channel and also $\textbf{h}_{ck}^{\dag} \cdot \textbf{W}_c$ from Charlie-Eve produce independent identically distributed $\mathcal{CN}$ random variables  with unitary variance \cite{Wang2020}. This enables the approximations $\left|\textbf{h}_{AE,k}^{\dag} (\textbf{h}_a/\|\textbf{h}_a\|)\right|^2 {\sim \;} \exp(1)$, $\|\textbf{h}_{AE,k}^{\dag} \textbf{W}_a\|^2 \sim \text{Gamma}(N_A - 1, 1)$ and $\|\textbf{h}_{c,k}^{\dag} \: \textbf{W}_c\|^2 \sim \text{Gamma}(N_C - 1 , 1)$.

\subsection{Performance metric}

Considering that Alice transmits codewords at a rate $R_b$ with a secrecy rate $R_S \leq C_S$, the redundancy rate can be defined as $R_e = R_b - R_S$. Then a secrecy outage event occurs when the channel capacity of any Eve is higher than the redundancy rate that Alice can provide, i.e., $C_E > R_e$.

In a multiple passive Eves scenario, whose Channel State Information (CSI) are unknown, the secrecy performance is addressed in terms of the Secrecy Outage Probability (SOP), since the only available information about the Alice-Eve channel is its statistics. Thus, the SOP is defined as
\begin{equation}
    SOP = 1 - \Pr\left(\max_{k \in K} \gamma_{E,k} < \beta \right),
    \label{eq: SOP}
\end{equation}
\noindent which is the complement of the probability that the highest SIR among all Eves is less than the threshold $\beta = 2^{R_e} - 1$. This means that higher values of secrecy can be obtained by implementing the aforementioned PLS techniques to reduce $\gamma_{E,k}$ as much as possible.

\section{Numerical Results}
\label{section: Num_sims}

Various simulations with different parameters were performed to evaluate the relationship between the SOP and the decrease of the SIR for the $k$-th Eve. Since the V2X network model is randomly generated, the coordinates of each node and street change with each run. To provide more consistent results, the curves presented below are the average of multiple realizations of each simulation configuration. 

\begin{figure}[hbt]
    \begin{center}
        \setlength{\unitlength}{0.0105in}
        \subfloat[Artificial Noise]{\includegraphics[width=.5\linewidth]{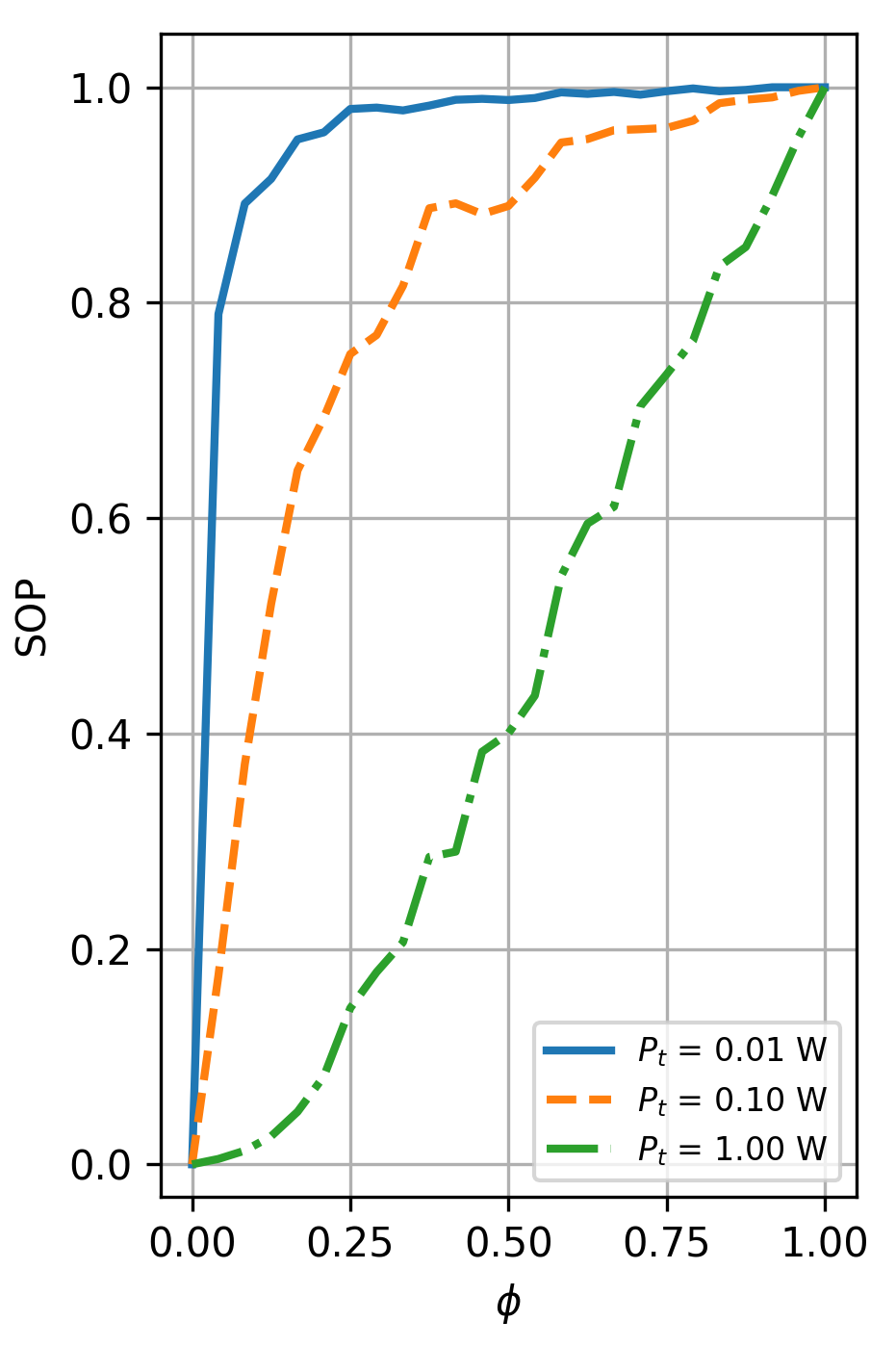}}
        \subfloat[Cooperative Jamming]{\includegraphics[width=.5\linewidth]{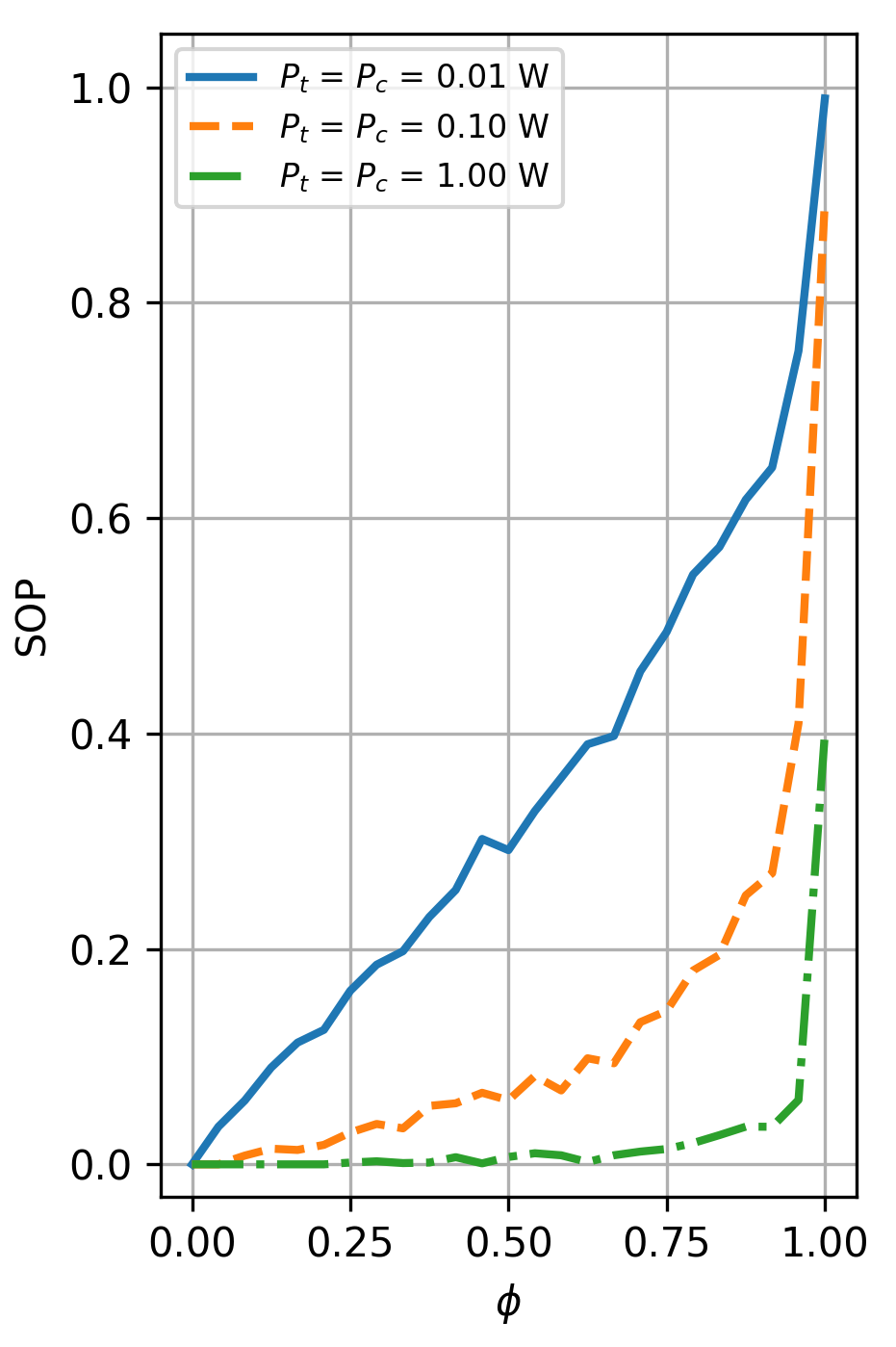}}
    \end{center}
    \caption{\label{fig: SOP_Pots} SOP versus $\phi$ (25 realizations) for the AN and CJ with different available power \{0.01, 0.1, 1\} W. $\beta$ = 0 dB, $\alpha$ = 3, $N_A$ = $N_C$ = 4, $\lambda_E$ = $\lambda_C$ = $10^{-6}$/$m^2$ , $\mu_E$ = $\mu_C$ = $10^{-3}$/m, $r$ = 3 km.}
\end{figure}

Fig. \ref{fig: SOP_Pots} illustrates the SOP for different $P_t$ and $P_c$ values, ranging from 10 mW (10 dBm) to 1 W (30 dBm). As expected, when the devices have more power available for interference, the SOP is greatly reduced. However, for the AN scenario secrecy is still not guaranteed when $\phi$ grows. For CJ, the SOP increases in a much slower rate due to the larger amount of nodes jamming the signal received by the Eves.

\begin{figure}[hbt]
    \begin{center}
        \setlength{\unitlength}{0.0105in}%
        \includegraphics[width=\linewidth]{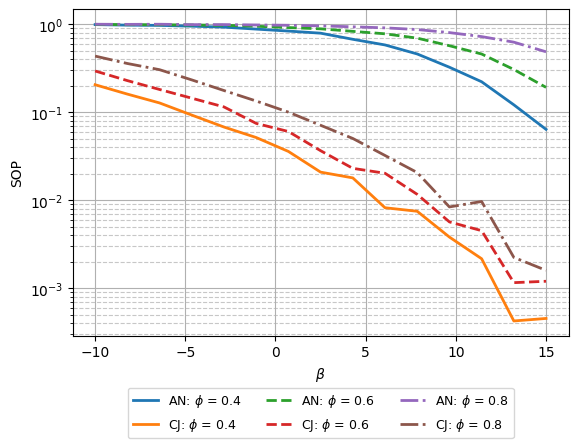}
    \end{center}
    \caption{\label{fig: SOP_Betas} SOP versus $\beta$ (50 realizations) for the AN and CJ with different power allocation ratios \{0.4, 0.6, 0.8\}. $\alpha$ = 3, $P_t = P_c$ = 20 dBm, $N_A$ = $N_C$ = 4, $\lambda_E$ = $\lambda_C$ = $10^{-6}$/$m^2$ , $\mu_E$ = $\mu_C$ = $10^{-3}$/m, $r$ = 3 km.}
\end{figure}

Through the simulation results presented in Fig. \ref{fig: SOP_Betas}, it can be easily noted that as $\beta$ increases the SOP decreases, because the criteria for secrecy failure is becoming more selective. Furthermore, $\phi$ have an opposing effect when compared to $\beta$, suggesting that for higher threshold values to guarantee low SOP, more power needs to be allocated to interference. Because of that, in applications where the devices have limited power (such as IoT and V2X), CJ is a more economic approach as long as there are sufficient nearby auxiliary nodes.

\begin{figure}[ht]
    \begin{center}
        \setlength{\unitlength}{0.0105in}%
        \includegraphics[width=\linewidth]{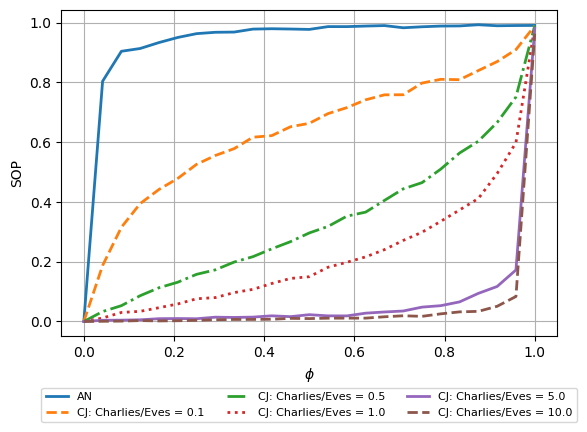}
    \end{center}
    \caption{\label{fig: SOP_CE_Ratios} SOP versus $\phi$ (25 realizations) for the AN and CJ with different $\lambda_C$/$\lambda_E$ ratios \{0.1, 0.5, 1, 5, 10\}. $\beta$ = 0 dB, $\alpha$ = 3, $P_t = P_c$ = 10 dBm, $N_A$ = $N_C$ = 4, $\lambda_E$ = $10^{-6}$/$m^2$ , $\mu_E$ = $10^{-3}$/m, $r$ = 3 km.}
\end{figure}

In Fig. \ref{fig: SOP_CE_Ratios}, it is evaluated the influence that the proportion of Charlies to Eves have on the SOP. This is achieved by implementing different values of intensities ($\lambda$ and $u$) for the Poisson processes that generate these nodes. The SOP grows rapidly in the AN, indicating that the available power is insufficient to guarantee secrecy with the given Eve density. For the CJ cases, however, as the number of Charlie nodes rises, the SOP starts to reduce, making the communication viable even for higher values of $\phi$. When there are more Charlies than Eves it is shown that very little power needs to be applied in each device to provide a low SOP.

\section{Conclusion}
\label{section: conclusion}

In this paper, a stochastic geometric approach was presented as a method to randomly generate V2X network models. The coordinates of these elements were then used to evaluate the effectiveness of PLS techniques in different realizations of vehicular networks subjected to path loss with NLoS. 

Both AN and CJ were introduced based on the analytical signals that the involved nodes transmit. Next, expressions were obtained for the SIR of Bob and the $k$-th Eve. Finally, the SOP was computed to evaluate the level of information security provided by the presented PLS techniques. 

Based on numerical results, it can be concluded that PLS can provide additional security for the V2X networks with relative low power cost, specially when both the techniques are combined. It is also noted that in the CJ scenario, when there are more Charlies in the proximity, the security increases. Therefore, the urban networks are the most benefited by this technique, since it is expected a higher density of wireless devices in the same area in these environments.

\end{document}